\documentclass[aoas]{imsart}

\usepackage{lscape, rotating}   
\usepackage{booktabs}   
\usepackage{textgreek}
\usepackage{ccaption}
\usepackage{ragged2e} 
\usepackage{mathtools, stackrel, enumerate}
\usepackage{tabularx}
\usepackage[stable]{footmisc}
\usepackage{floatrow}
\RequirePackage{epigraph}
\RequirePackage{amsthm,amsmath,amsfonts,amssymb}
\RequirePackage[numbers]{natbib}
\RequirePackage[colorlinks,citecolor=blue,urlcolor=blue]{hyperref}
\RequirePackage{graphicx, subcaption}

\startlocaldefs

\usepackage{xcolor}
\usepackage{sectsty} 
\sectionfont{\color[HTML]{FF3300}}
\subsectionfont{\color[HTML]{336666}}

\theoremstyle{plain}

\theoremstyle{remark}

\def\bx{\boldsymbol{x}}

\endlocaldefs

\setcitestyle{authoryear,round}
\begin{document}

\begin{frontmatter}


\textbf{\uppercase{Uniting Machine Intelligence, Brain and Behavioural Sciences\\ 
\smallskip
to Assist Criminal Justice}}

\runtitle{Machine Intelligence, brain and behavioural sciences, and law}

\thankstext{T1}{I thank Simon Chesterman, Martin F.,  Václav Janeček, Marcus R. Munafò,  Guy Nagels, Huy Phan, Sarah Rosanowski, Xiaojun Wang, and Bangdong Zhi for comments and criticisms on earlier versions of the paper. I thank a criminal judge (English law) and a retired civil lawyer (Belgium law) for helpful discussion and suggestions.}

\begin{aug}
\author[]{\fnms{Oliver Y.} \snm{Ch\'en}\ead[label=e1]{olivery.chen@bristol.ac.uk}}

\address{Faculty of Social Sciences and Law, University of Bristol\\
 \printead{e1}}
\end{aug}

\vspace{2cm}
\bigskip

\textcolor[HTML]{336666}{This is a working paper.\\ For comments, criticisms, or literature/case suggestions, please email me.\\ I will do my best to revise based on your suggestions and criticisms. \\
\vspace{5mm}
Last updated on \today.}

\vspace{2cm}

\begin{abstract}
I discuss here three important roles where machine intelligence, brain and behaviour studies together may facilitate criminal law. First, predictive modelling using brain and behaviour data may support legal investigations by predicting categorical, continuous, and longitudinal \textit{legal outcomes of interests} related to brain injury and mental illnesses. Second, psychological, psychiatric, and behavioural studies supported by machine learning algorithms may help predict human behaviour and actions, such as lies, biases, and visits to crime scenes. Third, machine learning models have been used to predict recidivism using clinical and criminal data whereas brain decoding is beginning to uncover one’s thoughts and intentions based on brain imaging data. Having dispensed with achievements and promises, I examine concerns regarding the accuracy, reliability, and reproducibility of the brain- and behaviour-based assessments in criminal law, as well as questions regarding data possession, ethics, free will (and automatism), privacy, and security. Further, I will discuss issues related to predictability \textit{vs.} explainability, population-level prediction \textit{vs.} personalised prediction, and predicting future actions, and outline three potential scenarios where brain and behaviour data may be used as court evidence. Taken together, brain and behaviour decoding in legal exploration and decision-making at present is promising but primitive. The derived evidence is limited and should not be used to generate definitive conclusions, although it can be potentially used in addition, or parallel, to existing evidence. Finally, I suggest that there needs to be (more precise) definitions and regulations regarding when and when not brain and behaviour data can be used in a predictive manner in legal cases.
\end{abstract}


\begin{keyword}
\kwd{Machine learning} 
\kwd{AI}
\kwd{brain science}
\kwd{behavioural science}
\kwd{predictive modeling}
\kwd{criminal justice}
\kwd{law}
\kwd{free will}
\kwd{automatism}
\end{keyword}

\end{frontmatter}

\newpage

\epigraph{
I am a brain, Watson.
The rest of me is a mere appendix.
Therefore, it is the brain I must consider.

\vspace{2mm}
\textit{The Adventure of the Mazarin Stone}
}

\section{Introduction}

\begin{figure}[H]
\includegraphics[width=110mm]{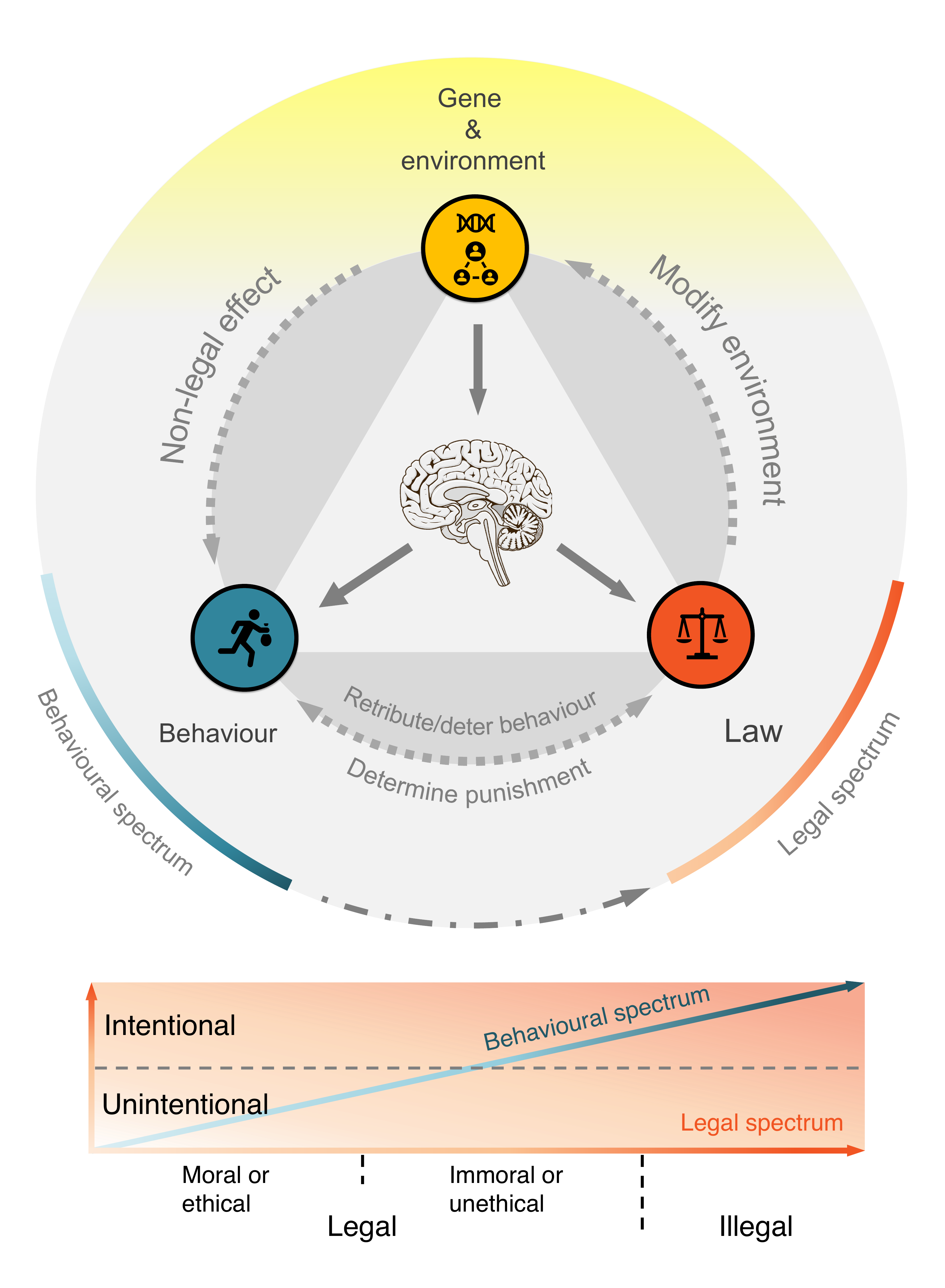}
\caption{The relationship between the brain, behaviour, and law. }
\medskip
\parbox[c]{\hsize} {\textbf{Top}: At the tips of the \textbf{triangle} are genetic and environmental factors, human behaviour, and law. The brain’s structures and functioning are determined by genetic and environmental factors and their interactions (top arrow). The brain dictates human behaviour (left arrow) and helps design and modify the law (right arrow). In the \textbf{inner circle}, genetic and environmental factors, along with the brain, affect human behaviour (left dashed arrow). (Criminal) behaviour determines reasonable punishment one receives from the law, and in turn, the law retributes criminal human behaviour and deters future crimes (bottom dashed arrow). Law governs and changes the environment in which one lives (right dashed arrow). On the \textbf{outer circle} are the behaviour spectrum and their corresponding legal spectrum. \textbf{Bottom}: Depending on the behaviour and intention, one receives different levels of legal judgment or punishment.
}
\label{Brain_law_triangle}
\end{figure}

All human laws are products of the brain. On the one hand, the laws are designed, defended, and delivered by involving functions of the specialised brain areas. On the other hand, understanding the functions and dysfunctions of the brain may assist jurors and judges in evaluating one’s thoughts, intentions, and actions and helping them to classify wrongdoings due to irregular brain activities, poor judgments, and criminal dispositions (see \textbf{Figure} \ref{Brain_law_triangle}). 

Recent development in machine intelligence, brain and behavioural sciences, and imaging technology have shown promises 
to use brain and behaviour signals to assess and predict various mental and cognitive faculties, such as intention \citep{haynes2007reading}, mental states \citep{reinen2018human}), psychological illness \citep{cao2018cerebello}, cognition \citep{chen2019resting}, as well as their behaviour outputs, such as telling lies \citep{farah2014functional}. The neurobehavioural and neuropsychological findings and their supporting technological frameworks, if proven broadly effective and reliable, may contribute to the jurisprudence in several ways: to separate false statements from facts, to exculpate those suffering from mental illness or insanity, to distinguish innocent from guilty, and to determine the degrees of fines, liabilities, or sentences.

In parallel, we have seen increasing usage of AI in the criminal justice system, such as ``the world's first robot lawyer'' \footnote{DoNotPay (\url{https://www.donotpay.com}).}, the consideration of AI-based legal solutions in Estonia \footnote{The Estonia Ministry of Justice clarifies that ``Estonia does not develop AI judge[s]''; rather, ``Ministry of Justice is looking for opportunities for optimization and automatization of court’s procedural steps in every types of procedures'' (\url{https://www.just.ee/en/news/estonia-does-not-develop-ai-judge}).}, robot mediators in Canada, and AI judges in Chinese courts \citep{chandran2022malaysia}. Meanwhile, recent years have seen a growing trend of multidisciplinary research intersecting machine intelligence, brain and behavioural sciences, and criminal justice (see \textbf{Figure} \ref{Lit_review}). 

Such adaptations, innovations, and expansions, however, have received mixed views. Noticeably, a recent case in Sabah, Malaysia, where a man was sentenced via the assistance of an artificial intelligence tool, has stirred legal outrage \citep{chandran2022malaysia}. The supporters argue that ``AI-based systems make sentencing more consistent and can clear case backlogs quickly and cheaply, helping all parties in legal proceedings to avoid lengthy, expensive and stressful litigation''. Yet the critics maintain that ``it is unconstitutional'', increases bias, and that, compared to human judges, AI does not consider mitigating factors or use discretion. In the middle ground are opinions suggesting that some decisions might properly be handed over to the machines, but transparency and explainability \citep{chesterman2021through} as well as regulations (in the sense of public control \citep{chesterman2021we}) are needed.

The thesis of uniting machine intelligence, brain and behaviour science to assist criminal justice, therefore, must confront three general challenges: technological difficulties, legal obstacles, and ethical concerns. For example, technologically, how reliable are predictive algorithms developed in labs when applied to real-world legal practices? Legally, even for reliable models, how should the legal profession adopt them to gather and deliver evidence during discussions and investigations? Compounding the technological and legal challenges, how can one prevent brain- and behaviour-based predictions from providing undue exculpatory evidence? Ethically, who owns our brain data; who determines whether one’s brain data can be recorded and analysed; and who judges whether the results can be used in a justice system?

\begin{figure}[H]
\includegraphics[width=130mm]{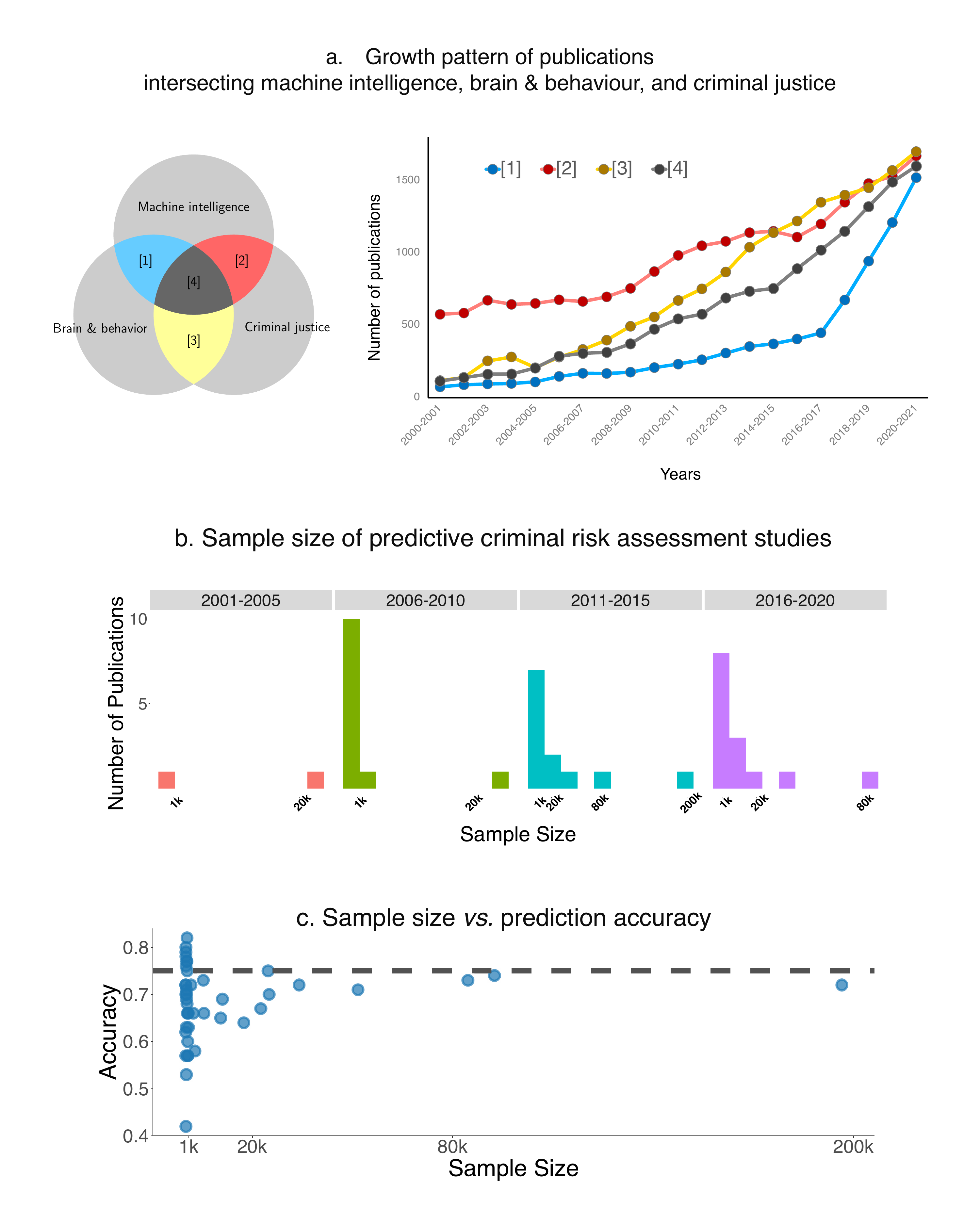}
\caption{The growing literature intersecting machine intelligence, brain and behavioural sciences and law.}
\smallskip
Caption on the next page.
\label{Lit_review}
\end{figure}

\begin{figure*}[h!]
  \contcaption{
  Continued.
  (a) \textbf{Publications whose texts contain key words related to machine intelligence, brain and behavoural sciences, and/or criminal justice.} We first define three sets of key words. Machine intelligence (MI) = $ \{ \textit{machine learning; artificial intelligence; machine intelligence} \}$; Brain and behaviour science (BB) = $ \{ \textit{brain science; neuroscience; behavioral science; behavioural science}  \}$; Criminal justice (CJ) = $\{ \textit{criminal law; law; criminal justice; criminal justice system; law; legal} \}$. We then define publications in [1]-[4] as $MI \cap BB \setminus CJ$, $MI \cap CJ \setminus BB$, $BB \cap CJ \setminus MI$, and $MI \cap BB \cap CJ$, respectively. The search is done using the \textit{advanced search} function provided by \textit{Google Scholar}, where, for example, the search query $MI \cap BB \setminus CJ$ is equivalent to entering \textit{``machine learning; artificial intelligence; machine intelligence; brain science; neuroscience; behavioral science; behavioural science -criminal -law; -law; -criminal -justice; -criminal -justice -system; -law; -legal''}. Note that a search for articles whose titles contain key words from all three sets (\textit{i.e.}, at least one key word from each of the three sets) yields no results. Note also that including a key word ``AI'' during the search would greatly reduce the search results, and, therefore, I have used ``artificial intelligence'' during the search. \textbf{(b) The trend of sample size in predictive criminal risk assessment studies.} Noticeably, there is a growing trend of larger samples investigated in the past decade than the one before. \textbf{(c) The relationship between sample size and prediction accuracy.} Large sample size does not necessarily endorse high prediction, although it may increase statistical significance (given the same effect size). But most predictions with high accuracy ($AUC > 0.7$) are from small sample size studies. Figure (a) is based on \textit{Google Scholar} data; figures (b) and (c) are produced based on data from \citep{fazel2022predictive}. 
  }
\end{figure*}

Inspired by these questions and concerns, here I present the promises, challenges, potential solutions and hopeful future directions of the brain- and behaviour-based assessment, prediction, and decision-making in jurisprudence. First, I argue that machine-assisted forensic science (see definition below) may facilitate legal investigations in three directions: (1) brain data-based mental faculty and behaviour prediction; (2) lie, intention, and crime scene detection; (3) general brain decoding. Next, I discuss accuracy, privacy, and ethical challenges when employing brain- and behaviour-based decision-making in legal practices. I argue that evidence derived from neural and behavioural analyses can presently provide supportive but not conclusive information. Finally, I argue that there needs to be clear definitions and regulations regarding when and when not brain and behaviour data can be used in a predictive manner in legal cases.

\textbf{Remark} 1. To facilitate discussion, throughout I consider brain and behaviour data broadly. For example, I define brain data as measurements of brain signals obtained either invasively or noninvasively (such as the action potentials, BOLD fMRI \footnote{Blood-oxygen-level-dependent (BOLD) functional magnetic resonance imaging (fMRI).} or electroencephalogram (EEG) recordings) \footnote{For brain decoding, it is at present perhaps more suitable and practical to use noninvasive measurements.}. I define behaviour data as the general outputs of the brain that may be useful in courts, such as measurements and recordings of language and hand movement. I will also consider intermediate, intangible brain concepts that mediate \footnote{Such mediation pathways can be further classified hierarchically. For example, brain data $\rightarrow$ thoughts $\rightarrow$ emotions $\rightarrow$ actions. } brain activities and actions, such as thought and intention. They are the outcome of the brain and simultaneously the potential cause of actions.

\textbf{Remark} 2. Many points I discuss in this paper may be primitive or controversial. I present them regardless for two reasons. First, I outline the possibilities of using AI/machine learning and brain/behaviour research to facilitate law (and \textit{vice versa}) and provide examples, concepts, and arguments to illustrate them. Some examples and arguments may be nascent or unusual to the norm; I do not intend to and cannot settle the linkages between AI/machine learning and the brain and behaviour and how they may evolve or advance the law in the future. A presentation of possibilities may help the readers conceptually, and I feel it is perhaps useful to expose these points as early as we can and let future research and practice contest, verify, and perhaps settle them. Second, I hope my presentations, potentially still in their infancy and contentious, may sprawl further discussions, either to modify my views or to expand my arguments.

\section{The roads taken: behavioural and neural approaches in assisting criminal justice}

\subsection{The road connecting behavioural data \footnote{More precisely, physiological indices or physiological signatures.} and deception via polygraph \footnote{I credit many cases to \citep{synnott2015review} and \citep{NAS2003polygraph}; see them for thorough reviews regarding the history of polygraph and the validity of it, respectively.}}

\label{sub: road_from_behaviour}

\underline{Prelude}. Italian criminologist and physician Cesare Lombroso proposed, rebutted and criticized, criminal atavism and biological determinism \citep{lombroso1911crime}, and created a device called ``Lombroso's Glove'', with which a subject could insert one's hand through a rubber membrane into a container of water to measure the volume of blood flow; a drop in blood pressure would suggest that the subject was lying \citep{kelly2004truth}. In 1904, Austrian-Italian psychologist Vittorio Benussi invented ``one of the first methods of lie detection, based on the ratio of inspiration to expiration length'' (p.36, \cite{zusne1984biographical}). In 1906, British heart surgeon James Mackenzie developed the Mackenzie-Lewis polygraph, an ink-writing polygraph, recording the pulse on a long roll of paper to detect problems such as an irregular heartbeat \citep{science_museum_london}. In 1915, American psychologist William M. Marston created the systolic blood pressure test \citep{marston1917systolic}, which became a component of the first polygraph (see below).

\underline{The first polygraph}. In 1921, American police officer John A. Larson invented the first polygraph, a device he called the ``cardio-pneumo psychogram'' which measured \textit{continuous} changes in blood pressure, heart rate, and respiration rate, and used it in criminal investigations \citep{segrave2003lie}. In 1923, Marston tried to admit the results of a polygraph test as evidence; this was rejected \citep{synnott2015review}. The reason for the rejection \footnote{``[W]hile the courts will go a long way in admitting experimental testimony deduced from a well recognised scientific principle or discovery, the thing from which the deduction is made must be sufficiently established to have gained general acceptance in the particular field in which it belongs'' \citep{fryeh1923}.} has become the so-called \textit{Frye Standard}, until it was replaced by the \textit{Daubert Standard} in 1993 (see below).

\underline{The first polygraph testing procedure}. In 1930, Leonarde Keeler (Larson's protégée), created arguably the first polygraph testing procedure \citep{synnott2015review}, the Relevant/Irrelevant Question Technique \citep{keeler1930deception}. Keeler also made the polygraph portable with simultaneously recorded pulse rate, blood volume change, and breathing, with a later addition of galvanic skin response (GSR) channel in 1939 \citep{grubin2005lie} His device was purchased by the Federal Bureau of Investigation (FBI) and became the prototype for the modern polygraph \citep{sullivan2002concise}. Soon, other polygraphs emerged, noticeably the Berkeley Psychograph in 1936 (the major difference between the Berkeley Psychograph and Keeler's design was the inclusion of a new design in the pulse-blood pressure unit \citep{berkeley_psychograph}), Darrow Behar's Behavior Research Photopolygraph in 1932 \citep{darrow1932behavior}, and John E. Reid's device which recorded both muscular activity and blood pressure changes in 1945 \citep{reid1945simulated}.

\underline{After the Second World War}. Reid developed the Comparative Question Test (CQT) \citep{reid1947revised}, replacing Keeler’s Relevant/Irrelevant Question technique, and has become the most widely used technique until today \citep{synnott2015review}. Alternative polygraph methods have also been developed during that period of time, including Orienting Response Theory \citep{sokolov1963perception}, the Guilty Knowledge Test (GKT), also known as the Concealed Information Test (CIT \citep{maclaren2001quantitative}. Since 1993, the \textit{Daubert Standard} \citep{daubert1993} has gradually replaced the \textit{Frye Standard} for the admissibility of expert testimony, whereby there is no longer a need for scientific evidence to have garnered the so-called ``general acceptance'' in courts by the scientific community to be considered admissible \citep{giannelli1997polygraph}.

\underline{Debates and modern usages}. The history of the debates about the accuracy of polygraph is perhaps as long as the history of polygraph. On the one hand, the American Polygraph Association ``claims that the polygraph has a high degree of accuracy in detecting truthfulness or deception, with research studies published since 1980 reporting average accuracy rates ranging from 80 to 98 percent'' \citep{NAS2003polygraph}. On the other hand, the US Committee on Government Operations concluded in 1965 that ``[t]here is no lie detector, neither man nor machine. People have been deceived by a myth that a metal box in the hands of an investigator can detect truth or falsehood'' \citep{united1965use}, when a proposal was made to use the polygraph to screen federal employees. Additionally, Honts, Kircher, and Iacono raised the issues of countermeasures \citep{honts1994mental, iacono1992use}. From a relatively neutral standpoint, Kleinmuntz and Szucko called for better research practices and methodologies \citep{kleinmuntz1984lie}. 

At present, the United States carries out thousands of polygraph tests each year on job applicants and current employees through a variety of agencies \citep{NAS2003polygraph}. At present, 19 States admit polygraph results under stipulation by the parties; the State of New Mexico, under strict evidentiary rules, permits polygraph results into evidence without a stipulation \citep{matte_legal}. In Canada, the Supreme Court rejected the use of polygraph results as evidence in court \citep{r1987}; but the polygraph results are admissible ``in civil and labor courts at the discretion of the judge'' \citep{matte_legal}. In England and Wales, polygraph tests is an option to monitor serious sex offenders on parole \citep{offender2007} and have, since 2014, become compulsory on high-risk sex offenders \citep{bowcott2014lie}. In Belgium, over 300 CQTs are performed annually \citep{meijer2004lie} and the Supreme Court ``afforded polygraph results judicial notice of acceptance with certain requirements that assure the reliability of the test and protect the rights of the defendant/examinee'' \citep{matte_legal}. In Poland, the Supreme Court is against the use of a polygraph during a hearing, except (only) during criminal proceedings, where there has already been pending charges against a person (\textit{i.e.}, after presenting the charges) and only by a court expert and only with the consent of the accused \citep{rudak2015supreme}.

\underline{A comprehensive review regarding the scientific evidence on the polygraph}. The National Research Council formed a Committee to review the scientific evidence on the polygraph \citep{NAS2003polygraph}. Based on 52 sets of subjects in the 50 research reports of studies (including 3,099 polygraph examinations) conducted in a controlled laboratory testing environment, the Committee concluded that ``in populations of examinees such as those represented in the polygraph research literature, untrained in countermeasures, specific-incident polygraph tests for event-specific investigations can discriminate lying from truth telling at rates well above chance, though well below perfection'' and that the variability of accuracy across studies is high; they also cautioned that the scientifically acceptable laboratory and field studies (with a midrange between 0.81 and 0.91) ``most likely over-states true polygraph accuracy in field settings involving specific-incident investigations''. Finally, the Committee suggested that ``[p]olygraph examinations may have utility to the extent that they can elicit admissions and confessions, deter undesired activity, and instill public confidence, although such utility is separate from polygraph validity" \citep{NAS2003polygraph}.

\subsection{The road connecting brain abnormality and conviction reduction via legal arguments \footnote{See \citep{greely2019neuroscience} for a thorough review.}}
\label{sub: road_from_the_brain}

One way to apply brain science in law is to reduce sentences by showing that the brain of the convicted is not functioning to its normal capacity, due, for example, to brain damages and mental illnesses. A concrete example where brain science has directly assisted criminal law was the Sheila Berry case \citep{commonwealth2014berry}. In brief, the initial conviction of murder in the first degree (on a theory of extreme atrocity or cruelty) was reduced to murder in the second degree because of Berry's lack of criminal responsibility: due to a combination of a damage to the frontal lobe of her brain (at the age of fourteen) with subsequent mood and personality changes, a prolonged history of mental illness afterwards (including bipolar diagnosis at age eighteen), and a brain tumor in her cerebellum (found at age twenty and doubled in size four years later - three years prior to the murder).

Another way to apply brain science in law is to argue for failing to introduce evidence of relevant brain damage and cognitive disorders into the sentence proceeding \citep{strickland1984}. For example, Jerry Ray Davidson's death sentence was vacated, after he petitioned for a post-conviction relief wherein he argued receiving ineffective assistance of counsel at trial \citep{davidson2014}, by the Tennessee Supreme Court which found his trial counsel failed to ``give the jury any mitigating information regarding [his] intellectual and cognitive deficits'' \citep{greely2019neuroscience}.

A third way is to use brain science to evaluate whether one is competent to ``understand the criminal charges against them, appreciate the consequences of a trial, including criminal punishments at stake, are able to communicate and assist their attorneys in their defense, and understand the nature of the proceedings against them'' \citep{dusky1960} or stand the trial  (due to an inability to assist in one's own defense due to, for example, addiction to medications or potential brain damage) \citep{US2011}, and proceedings are suspended until/if a defendant's competency has been restored \citep{greely2019neuroscience}.  

Finally, and perhaps more generally, brain science may assist law by examining free will; more specifically, by distinguishing criminal behaviour due to free will from criminal behaviour due to brain damage (which disrupts free will) (see \textbf{Section} \ref{subsection: free_will} for details).

\subsection{Converging roads?} \label{sub: converging_two_roads}

Despite advances, the practices of using behavioural data for detecting deception, and using evidence of brain or mental illness to assistant criminal defendants have so far been largely done in isolation. Naturally, there are some gaps and potential unexplored intersections, and therefore, needs to bridge them. \textbf{First}, although brain or mental illnesses, once confirmed, can help to argue for a potential reduction of sentence, it remains to show how to (reliably) identify, assess and predict brain and mental illnesses. In \textbf{Section} \ref{section: predicting_brain_injuries_mental_illness_behaviour}, I will discuss the brain areas associated with control of impulsiveness, emotions, judgments, and violent and criminal behaviour, and define three types of outcomes of legal interest for which one aim to predict via a model. \textbf{Second}, although it is possible to detect lies via polygraphs using behavioural data, it remains to show whether this is possible or can be improved using brain data alone, or by combining behavioural and brain data, and to expand lie detection to detecting bias and visits to crime scenes. I will discuss this in \textbf{Section} \ref{section:lies_bias_crime_scenes}. \textbf{Third}, while we have seen brain and behavioral decoding in specific areas (\textit{e.g.}, lie detection), would it be possible to perform brain and behavioral decoding more generally; in other words, can one use brain and behaviour data to predict different types of legal outcomes of interest? And are such general predictions subject to free will and automatism? I will attempt to take these issues in \textbf{Section} \ref{sec:general_prediction}. \textbf{Finally}, I will discuss the challenges of merging these two roads in \textbf{Section} \ref{section:challenges}, including privacy, ethics, and prosecutory abuse (\textbf{Section} \ref{subsection: privacy_ethics_prosecutorial_abuse}), accuracy, reliability, and reproducibility (\textbf{Section} \ref{subsection: accuracy_reliability_reproducibility}), the challenges of predictig future actions (\textbf{Section} \ref{subsection: predicting_future_actions}), the issue of population-level prediction \textit{vis-à-vis} personalized prediction (\textbf{Section} \ref{subsection:population_vs_personalized}), the choice between prediction accuracy and explainability (\textbf{Section} \ref{subsection: predictability_vs_explainability}), and how to mitigate or address some of these challenges. 

One way to link these seemingly largely isolated sections is to refer to the triad of brain, behaviour and law in \textbf{Figure} \ref{Brain_law_triangle}. Polygraphs establish the link from behaviour inputs to law. Criminal reduction based on mental illness in \textbf{Section} \ref{sub: road_from_the_brain} links brain and law but only if the illness is proven; \textbf{Section} \ref{section: predicting_brain_injuries_mental_illness_behaviour} aims at linking brain data to brain illnesses via machine intelligence. \textbf{Section} \ref{section:lies_bias_crime_scenes} links brain data to behaviour via machine intelligence. \textbf{Section} \ref{sec:general_prediction} links both brain and behavoiur data to law via machine intelligence and extend the aforementioned links to general cases. \textbf{Section} \ref{section:challenges} discusses the challenges of establishing these links and potential solutions. Noticeably, there is a genetic and environmental component in \textbf{Figure} \ref{Brain_law_triangle}. While there are indeed genetic underpinnings of brain illness (such as Parkinson's disease), I shall leave the discussion of genetic factors in criminal justice to those who are more capable. I will, however, briefly discuss the environmental and social factors in \textbf{Section} \ref{sec:final_remarks}.

\section{On predicting brain injuries and mental illnesses} \label{section: predicting_brain_injuries_mental_illness_behaviour}

Brain and behaviour studies offer promises to advance \textbf{machine-assisted forensic science}. Here, by machine-assisted forensic science, I mean the practice of using AI and machine learning algorithms to derive evidence from biological data to support decision-making in legal investigations. Our consideration of the biological data goes beyond fingerprints and DNA, which are traditionally used in legal investigations; rather, they extend to include data recorded from the brain and human behaviour through brain imaging, sensors, or wearable computers. 

The disruption of the functioning and structure of the human brain may result in unusual behaviours. For example, the prefrontal cortex (PFC) is involved in executive function, cognition, control of impulsive behaviour, emotional regulation, judgment (and moral reasoning), organisation, planning, and decision-making \citep{stuss1984neuropsychological}. When one’s PFC is damaged, one may lose, in part or entirety, morality and propriety \citep{sapolsky2004frontal} and present behavioural disinhibition and impaired intellectual faculty \citep{reber2019frontal}. The studies on Phineas Gage suggested the link between frontal lobe damage and radical behaviour changes. Since then, the PFC lesions and damage have been shown to be associated with violent and criminal behaviour \citep{brower2001neuropsychiatry}. Besides the PFC, impairment of the amygdala may also be related to irregular emotion, decision-making, and social judgment \citep{anderson2001lesions, gupta2011amygdala, phelps2002cognitive}. Recent works on functional and structural MRI have additionally shown that offenders with relatively low anterior cingulate (ACC, a limbic area associated with error processing, conflict monitoring, response selection, and avoidance learning) activity are more likely to be rearrested \citep{aharoni2013neuroprediction}. 

With recent development in predictive modelling, one can now link a specific brain region (\textit{e.g.}, the PFC) to an \textbf{outcome of legal interest}. Here, I define a multivariate, potentially high-dimensional input ($\bx$) and an outcome of legal interest ($y$) as follows.  

Let $\bx=(x_1,x_2,\ldots,x_p)$ be a $p$-dimensional feature variable consisting of $p$ features (\textit{e.g.},  brain activities from $p$ brain areas). One could extend $\bx$ to behaviour features, or features combining both behaviour and neural data. 

\begin{figure}[H]
\includegraphics[width=140mm]{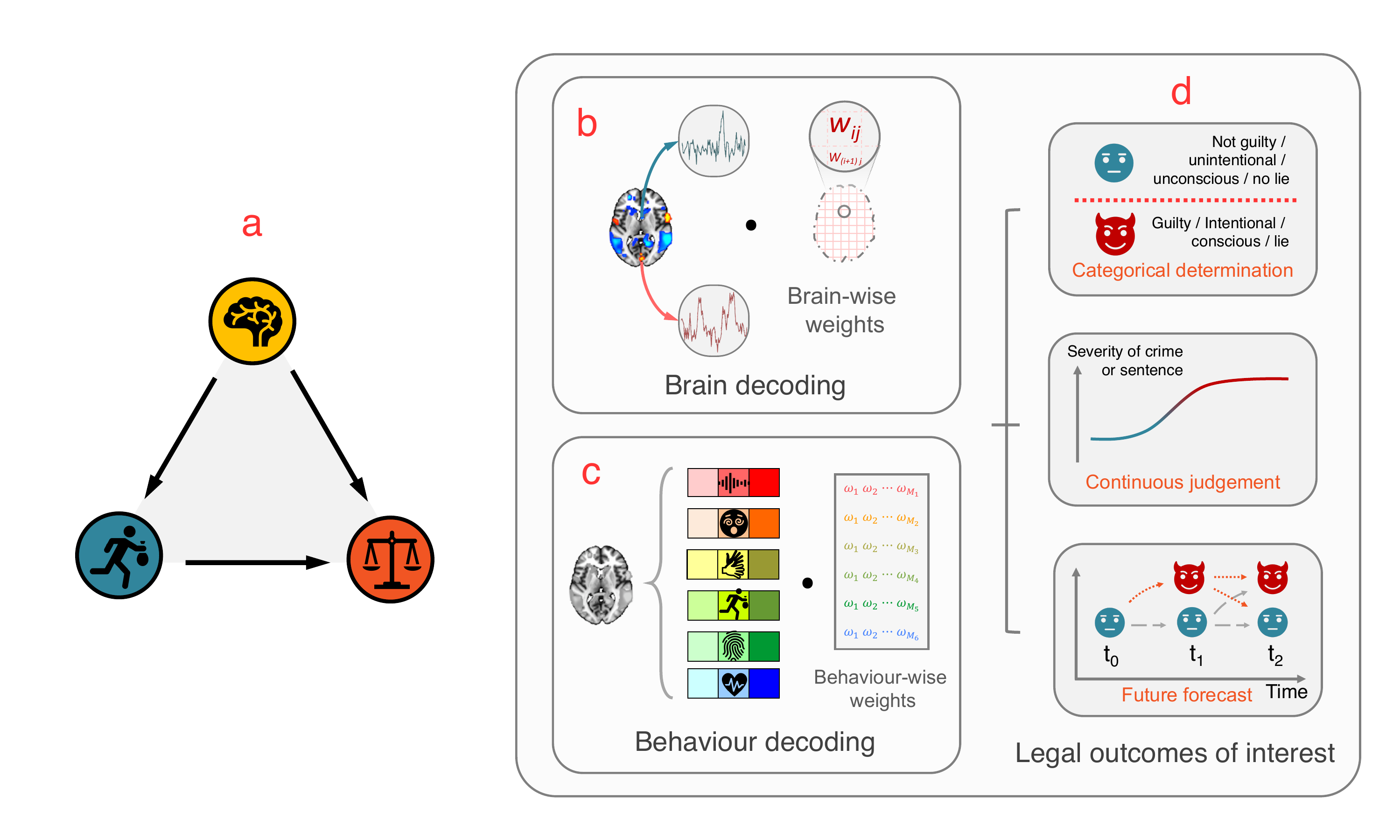}
\caption{A schematic representation showing how predictive models developed on brain and behaviour data may assist legal decision-making.}
\medskip
\parbox[c]{\hsize} { \textbf{(a)} The triangle between the brain, behaviour, and law. One can either use brain data directly to assist legal judgement; one can also use behaviour data to assist legal judgment (because behaviours are outputs of the brain). \textbf{(b)} Brain decoding. Individual brain data are coupled (\textit{e.g.}, via point-wise multiplication) with trained brain-wise weights to yield outcomes of legal interests. \textbf{(c)} Behaviour decoding. Individual behaviour data are coupled with trained behaviour-wise weights to yield outcomes of legal interests. \textbf{(d)} Three types of outcomes of legal interests. Top right: Categorical determination (\textit{e.g.}, whether someone acted intentionally or unintentionally). Middle right: Continuous judgment (\textit{e.g.}, how severe is the crime). Bottom right: Future forecast (\textit{e.g.}, what is the likelihood of someone committing a crime, or re-offend, or rehabilitate in the future?).
}
\label{Prediction}
\end{figure}

I define an outcome of legal interest ($y$) as a label that shows a particular characteristic of an individual under legal investigation. The outcomes of legal interest can be, in general, grouped into one of the three types: categorical outcomes, continuous values, and longitudinal estimates (see \textbf{Figure} \ref{Prediction} (d) for more details).

The first group of outcomes of legal interest is \textbf{categorical outcomes}. Examples of categorical outcomes of legal interest are: whether someone committed a crime intentionally or not, whether someone is guilty or not, and whether one’s actions are punishable or not? Because the outcomes can take one of the defined categories (\textit{e.g.}, intentional \textit{vs.} unintentional), they are categorical. Related to categorical outcomes of legal interest within criminal law are three types of \textit{mens rea}: intention, recklessness, and negligence \footnote{\textit{Intention}: A defendant will be found to have intended a consequence if they desire the consequence to follow their actions). \textit{Recklessness}: A defendant was aware that a risk existed and went on unreasonably to take that risk). \textit{Negligence}: A defendant does not reach the standard of care any reasonable individual would take; by failing to do so, it caused harm(s) to the victim.}. Using machine learning algorithms, one can map individual brain and behaviour data onto individual categorical outcomes to classify subjects into different groups (say, whether the individuals fall into the group with an intention to kill); a trained model can then be used to predict intentions in new subjects given their brain and behaviour data.

If someone intends to kill another person and kills him/her, this is viewed more harshly than if someone is reckless as to whether an act they are doing might kill someone and then it (\textit{i.e.}, the action) kills them. To assess the severity of the outcome, one needs perhaps more refined outcomes of legal interest, that is, \textbf{continuous outcomes}. They are continuous because the outcomes can potentially take any values within a range (rather than discrete values). For example, given the brain and behaviour data, how severe (say, between 0 and 100) is one’s action; how likely (say, between 0 and 1) will they yield criminal actions? 

The third group is \textbf{longitudinal monitoring and forecasting}. For example, following the brain and behaviour data over time, what is the likelihood of someone committing a crime at any given point in time (monitoring), and in the future (forecasting)? Whereas the first two types of outcomes of legal interest are mostly trained and tested on observed data, the forecasting aspect deals with future data that are not as-of-yet observed. One, therefore, has to be extremely careful about predicting a person’s future intentions and actions \footnote{If a person has not committed any criminal acts yet (despite their mindset), it would be extremely onerous to suggest they should be punished or have their liberty restricted in any way due to thoughts that go through their mind on which they have not acted. Otherwise, this would be reminiscent of the \textit{Precrime} in Philip K. Dick's \textit{The Minority Report} and the \textit{Thought Police} in George Orwell’s \textit{Nineteen Eighty-Four}, who punishes thoughtcrime. Another related concepts are the \textit{Espers} (telepathic humans) in Alfred Bester's \textit{The Demolished Man}. 

On the other hand, if one could legally gather the data (\textit{i.e.}, in line with the GDPR), it could be helpful to assist with knowing which persons of interest in the community are so they can be surveilled, like the ‘heat list’ in the US. MI6 has a similar list in the UK. See \citep{fazel2019prediction} for a study on the prediction of violent reoffending in prisoners and individuals on probation.
} (see \textbf{Section} \ref{section:challenges}).

Returning to the general predictive framework, given individual brain/behaviour data $\bx$ and labelled outcome $y$, one can use AI/machine learning algorithms to develop (via training and test and cross-validation) a model $M: M(\bx) \rightarrow y$ that maps brain/behaviour data $\bx$ onto outcome $y$. Moreover, one can use this trained model $M$ to assess outcomes given data from a new subject $\bx^{new}$; in other words, $y^{new}=M(\bx^{new})$. If the model captures the relationship between brain and behaviour data and outcomes in a general population, and the data are of good quality, then the estimated outcome $y^{new}$ is expected to be close to its unobserved (but true) outcome. This may help assess potential outcomes during legal investigations when the outcomes are not directly/yet available.

The outcome of legal interest is not restricted to univariate cases (\textit{i.e.}, with only a single outcome). By evaluating the brain patterns across subjects, machine learning algorithms can identify neural markers related to several outcomes of legal interest. When brain data are unavailable, behavioural data (such as speech patterns) recorded semi-continuously, for example, from wearable computers, can serve as surrogate behavioural markers. 

Recently, using these technologies, one has begun to assess and predict abnormal mental states \citep{reinen2018human}, psychological episodes \citep{cao2018cerebello}, and the presence of dementia \citep{teipel2008novel} in neuroscience and psychology. These explorations have proved the concept of and provided technical foundations for making categorical (present of illness or not) or continuous (how severe the symptoms are) enquiries to determine insanity, culpability, or ability to form intent. In principle, too, these technological advances and neurobiological findings can be translated to criminal law, especially in the terrain of outcome prediction. 

But little do we know to what extent they can be applied and adopted in criminal law practice. A beginning can perhaps be made by testing these approaches using existing data that have been labelled (\textit{i.e.}, the outcome such as intention to kill has been confirmed) and applying these approaches in several areas in criminal law to (a) find which types of data may yield accurate performance; (b) pinpoint which sub-fields of criminal law may at present benefit the most from such explorations; (c) identify areas where the models are not performing well and investigate the gaps. Another attempt may be to discuss and explore the possibilities of performing predictions using ensembled data (in other words, combining both brain/behaviour information and traditional biological data such as fingerprinting and DNA in criminal law) (see \textbf{Section} \ref{section:lies_bias_crime_scenes}).

\section{On predicting human behaviour and actions} \label{section:lies_bias_crime_scenes}

Central to robust and fair jurisprudence is accurately and consistently distinguishing punishable defendants from innocent ones. Importantly, for human judges, this requires reliable and reproducible methods to separate, in terms of closed-form (yes or no) answers, truthful statements from partially true or completely wrong statements given by the defendants. 
 
In cross-examination, the counsel questioning the witness must ask ‘closed’ questions (\textit{i.e.}, questions that will only elicit a yes or no response). For example, “you were in Howard House on the night of Friday, 26 November; please answer only yes or not?” Oftentimes, however, when faced with mounting legal documents and testimonies, it may be challenging to ascertain the trustfulness and unbiasedness of the evidence/statements. Machine learning may help to perform data reduction and select features to make dichotomous predictions. 

One interfacing area between machine learning and brain science that may contribute to criminal law is crime scene detection, J.D. Haynes and colleagues at the Berstein Center in Berlin have been working on projects involving participants touring various virtual reality (VR) houses. After viewing these houses, the participants’ brains are scanned. Early results show that it is possible to identify the houses one had before been to based on one's brain data. This shows the promise of using brain imaging techniques to determine whether a suspect had been to a crime scene \citep{smith2013reading}. 

Although such explorations (and the methodologies implemented) demonstrated the possibility of making dichotomous predictions using brain data, they are not without limitations. One difficulty is distinguishing whether the suspect had been to the crime scene while committing the crime or one was there accidentally. Another complication is determining the threshold of the prediction. Oftentimes, predictive algorithms for dichotomous outcomes yield a (predicted) score ranging from 0 (the suspect was predicted to be \textit{surely not there}) and 1 (the suspect was predicted to be \textit{surely there}). One could subsequently set a threshold, say, 0.6, such that an individual with a predicted outcome above the threshold is judged to be at the scene, and one with a predicted score below the threshold is judged not at the scene. It is, therefore, not only a prediction problem but also a legal (and perhaps a philosophical) one to determine a suitable threshold (as the threshold decides the sensitivity and specificity of the results). 

Naturally, one would ask, at what stage(s) of the legal investigations should machine intelligence, brain, and behaviour data-based assessment and prediction be (more) involved? My views are as follows. During evidence collection, one may employ them when the witness is being interviewed by the lawyers preparing their witness statement (\textit{i.e.}, weeks or months before the trial). Once in the courtroom, the witness will be reading the witness statement prepared by someone else, which itself may tamper with their brain signals (\textit{i.e.}, if something is not worded exactly how they would have done), or they will only be answering yes/no questions. We cannot see at present a clear scenario where a witness (except perhaps the defendant) would agree to have one’s personal data collected and processed, in the same vein that lawyers and jurors would not agree to this. More studies, discussions, and debates are needed regarding this (see \textbf{Section} \ref{subsection: when_can_be_used} for a further discussion).

Important to social interaction are trust and cooperation. They are integral elements in maintaining and facilitating effective and truthful communication between the defence lawyers and the accused, the plaintiff and the accusers, and judges and jurors. Some neuroscientists and neuroeconomists are trying to explain the neurological and behavioural underpinning of trust and cooperation. What has not yet been well explored, but is important to the law in general and criminal law in particular, is to assess and predict trust and cooperation using the brain data \citep{mccabe2001functional}. By directly examining the neural correlates of trust and cooperation, one may begin to find optimal strategies (and awards) needed to evaluate (and promote) trust (and cooperation) between different parties during legal investigations.

Brain signals are generally not subject to fabrication or distortion (but see \textbf{Section} \ref{subsection: accuracy_reliability_reproducibility}). Social pressure may prevent one from expressing unpopular or socially unacceptable views \citep{fazio1995variability, nosek2002two}. Brain data preserve objective measurements of one’s thoughts and may provide alternative views regarding a suspect than potentially biases from the jurors' \footnote{Here, I assume that juror’s brain data could also be analysed  – likely jurors would not consent to this. Further work may discuss the possibility and usefulness of it in the future and, if useful, what is needed to make it possible.} \citep{korn2012neurolaw} (see also the privacy and ethics concerns in \textbf{Section} \ref{subsection: privacy_ethics_prosecutorial_abuse}). Brain imaging data are also beginning to show promises to advance lie detection. For example, two companies, No Lie MRI and Cephos, are presently offering fMRI-based lie detection services in the criminal justice context \citep{farah2014functional} (see issues below). Taken together, coupled with brain and behaviour data, analytical tools may potentially help to identify and correct biases and false statements during evidence collection and court debates.

Despite promises, brain and behaviour data-based detections of a lie, bias, and visits to a crime scene at present face several major shortcomings. The \textbf{first} is regarding accuracy. Although impressive - given the mostly noninvasive nature and relatively low resolution of the data, the accuracy of image-based lie detection is not high \citep{farah2014functional}. This may present a major problem for a practical adaptation of using brain and behaviour data in the context of criminal law at present. To be convicted, the legal standard is that the judge/jury must consider the defendant committed the crime ``beyond a reasonable doubt''. This is a very high standard of certainty that the defendant committed the crime; if any doubt exists, the defendant must be acquitted. In fact, for a murder trial, for instance, the defendant will face losing their liberty for many years. Therefore, for machine intelligence-based evidence to be useful, it must be highly accurate. The \textbf{second} is regarding reproducibility. Although laboratory and field tests have shown that detection results based on brainwave (\textit{e.g.}, P300-MERMER) stimulated by words/pictures relevant to a crime scene are laudable \citep{farwell2012brain, farwell2001using}, it is possible to manipulate one’s thoughts to conceal information during brain data-based detection \citep{hsu2019effect}. In parallel to continuing to improve the identification accuracy by designing better predictive algorithms and developing improved data collection methods, one potential way to address these issues is to consider combining brain imaging data with other traditionally used physiological and behavioural data, such as one’s facial expression \citep{ekman1969nonverbal}, tone \citep{scherer1985vocal}, and content of speech \citep{vrij2008nonverbal} via data fusion and ensemble-learning to assess, for example, lies, during legal proceedings (rather than using an isolated type of data).

\section{On general brain and behaviour decoding} \label{sec:general_prediction}

As science, technology, and law interface and integrate, it is perhaps worthwhile to sail into the treacherous water to discuss how the brain decoding of general thoughts, intentions, \textit{mens rea}, and “automate” process \footnote{One performs an action that typically requires cognition without much thought after performing it many times \citep{heiner1983origin}.}, as well as behaviour-based prediction, may assist (or threaten) law and criminal justice \citep{vilares2017predicting}.

\begin{sidewaysfigure}[H]
\includegraphics[width=230mm]{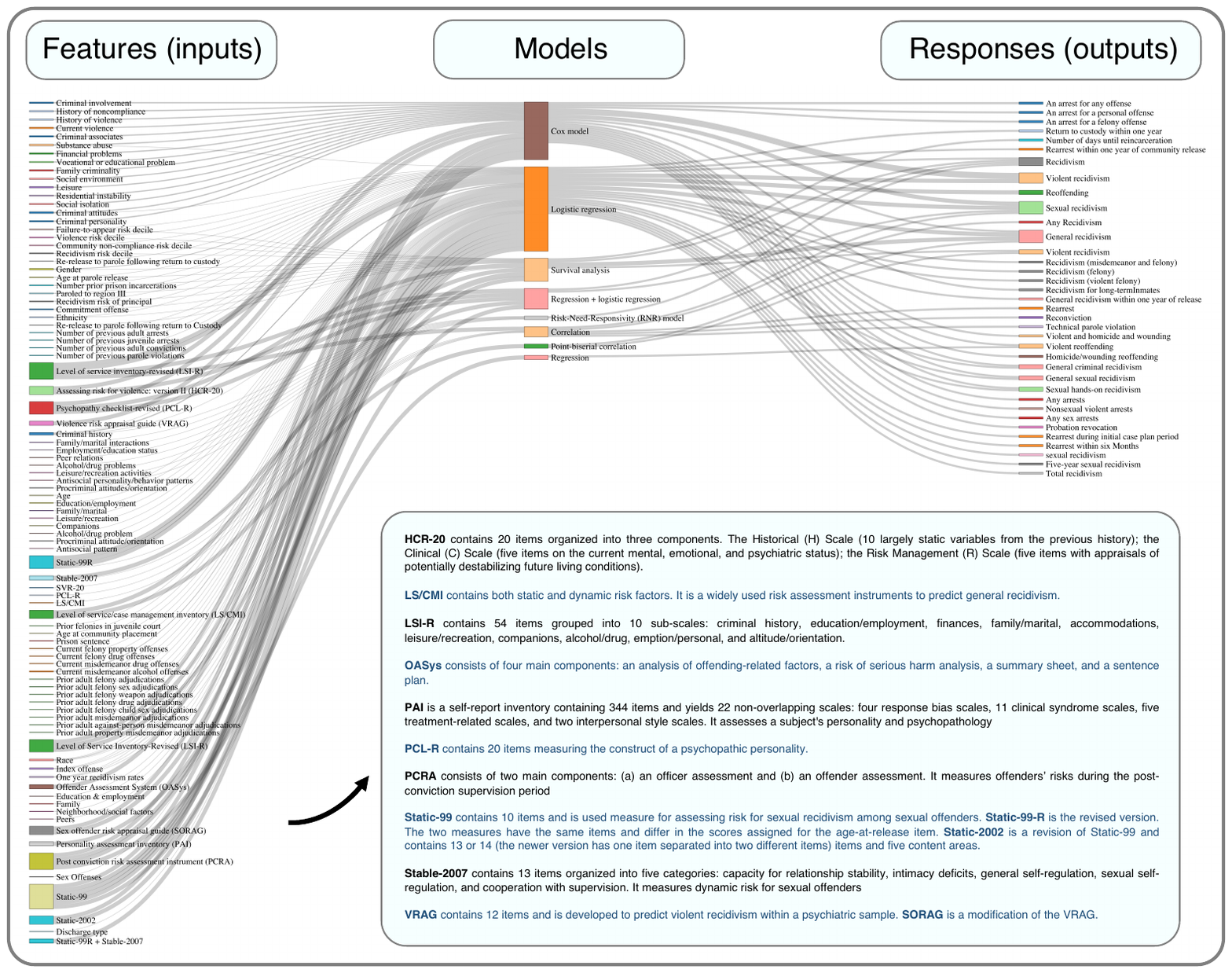}
\caption{Linking features and outcomes via models.}
\smallskip
\parbox[c]{\hsize} {
\begin{center}
Caption on the next page.
\end{center}
}
\label{feature_maps}
\end{sidewaysfigure}

\begin{figure*}[h!]
  \contcaption{
  Continued.
The features, methods, and predictable outcomes are identified from a subset of papers collected in a systematic review \citep{fazel2022predictive}. Specifically, I have only considered papers that report the features, methods, and outcomes \citep{brennan2009evaluating, farabee2010compas, fass2008lsi, dahle2006strengths, mills2007validity, dyck2018real, gordon2015evaluation, tsao2021exploratory, wormith2015predictive, barnoski2003washington, manchak2008utility, ostermann2013validating, vose2013predictive, howard2013identifying, lovins2018validating, harris2017field, rettenberger2017actuarial, walters2005use, cohen2018predicting, lowenkamp2013federal, luallen2016predictive, skeem2020using, boccaccini2017field, etzler2020dynamic, hanson2014field, martens2017predictive, smallbone2013short, veith2018predictive}. Note all papers considered contain an ROC analysis and report an AUC. Thus I have omitted ROC analysis in the figure. Additionally, I have excluded papers including only ROC and AUC, without using a specific (predictive) model. I only report a correlation analysis (between the features and outcomes) if it is the only ``model'' used (other than an ROC analysis) in a paper. In other words, if a paper has used the Cox proportional hazards model, correlation analysis, and reported the AUC, I only include the Cox model. The reason to discard correlation analysis is as follows. If variable $X$ is highly associated with $Y$, then $X$ is likely predictive of $Y$. But herein I consider prediction in the model (mapping) sense, $Y = \mathcal{M}(X)$, where $\mathcal{M}$ is a predictive model (map). 
}
\end{figure*}

\subsection{The need}
There is a need to extend brain and behaviour decoding to aspects beyond behaviour prediction and lie detection. Here, I define \textbf{brain and behaviour decoding} as using brain and behaviour data to assess \textit{general} outcomes of legal interest. 
Although lie and intention detection indeed fall into the general concept of brain decoding, their functions and goals are relatively specific (\textit{e.g.}, to tell whether the defendant is lying). Legal decision-making, however, is sometimes indirect and often requires assessment and predictions beyond lie and intention detections. For example, a murder charge can sometimes be reduced to a charge of manslaughter if the defendant lost control \citep{nicola2021use} \footnote{Note that \textit{provocation} is no longer a defence to murder in the UK (Coroners and Justice Act (2009, c.25)) \url{https://www.legislation.gov.uk/ukpga/2009/25/section/56}; but the \textit{loss of control} is still a defence to murder.}. Here, brain decoding may contribute to such a case by showing, via brain and behaviour analysis, that the defendant’s mental state has been \textit{continuously} disrupted (see \textbf{Section} \ref{subsection: predicting_future_actions} for further discussion).

\subsection{Technical promises} \label{subsection:legal_outcome_of_interest}

Experimental research in laboratories has demonstrated that it is possible to use brain data to unfold natural images and subjective contents we see \citep{naselaris2009bayesian, kamitani2005decoding}, movies scenes we watch \citep{nishimoto2011reconstructing}, hidden intentions we have \citep{haynes2007reading, haynes2011decoding}, and episodes of dreams we undergo \citep{horikawa2013neural}. These lines of evidence and methods suggest the possibility of potentially reconstructing (crime) scenes one had witnessed and wrongful actions one had taken by showing images of the scenes and actions and examining their corresponding neural activities. Additionally, by establishing these findings, they have provided pipelines to test these possibilities. 

Additionally, the ever-improving classification models, predictive algorithms, and longitudinal methods provide analytical foundations for assessing the three types of outcomes of legal interests in \textbf{Section} \ref{section: predicting_brain_injuries_mental_illness_behaviour}. Critically, two decades of research have shown that actuarial or hybrid clinical-actuarial systems, such as VRAG, Static-99, HCR-20, PCL-R (see \textbf{Figure} \ref{feature_maps}), developed using statistic methods connecting fixed factors and legal outcomes of interest (such as recidivism) perform better than clinical assessment and individualized assessments made by judges or jurors \citep{monahan2001rethinking, greely2019neuroscience}. In parallel, researchers have shown that functional and structural MRI scans (especially the anterior cingulate cortex (or ACC \citep{aharoni2013neuroprediction}) are potentially predictive of recidivism \citep{gaudet2016can}. 

The recent development of artificial neural networks, although relatively less explored in criminal justice, may provide novel insights into identifying complex, nonlinear hidden associations between brain and behaviour data and outcomes of the legal interest \citep{arrieta2020explainable}; also see predictability \textit{vs.} explainability in \textbf{Section} \ref{subsection: predictability_vs_explainability}.  

Yet, despite technical advances, most methods used in classification and prediction in criminal justices are still relatively simple. \textbf{Figure} \ref{feature_maps} has surveyed a few prominent methods aiming at training features (\textit{e.g.}, VRAG, Static-99, HCR-20, and PCL-R) to make criminal predictions of, for example, recidivism and rearrest. One notices that a large proportion of the works used Cox (proportional hazard) model, linear regression, logistic regression, and correlation analysis. Whereas I believe that if a simple model can address a complex question, it is a useful model (see also Occam's razor), and the surveyed methods in \textbf{Figure} \ref{feature_maps} have demonstrated the possibility of prediction and unequivocally advanced our understanding of the linkage between the input features and legal outcomes of interest. Yet, the prediction accuracy is relatively low (see \textbf{Figure} \ref{Lit_review}). Given a wealth of recently developed, more advanced computational, machine learning approaches (see above), it is perhaps useful to test and adopt some of these methods (\textit{e.g.}, a support vector machine with customized kernels may better distinguish those with a tendency to recidivism from those who do not when the two groups are not linearly separable;  a convolutional neural network may identify hidden patterns between VRAG, Static-99, HCR-20, and PCL-R and recidivism/rearrest) to, on the one hand, examine the utility of advanced models in legal predictions, and, on the other hand, potentially improve the prediction accuracy of current models (see also \textbf{Section} \ref{subsection: predictability_vs_explainability} for predictability \textit{vs}. explainability). A beginning can perhaps be made by compare these ``simple'' methods with advanced approaches on the same data in \textbf{Figure} \ref{feature_maps}.

\subsection{On free will and automatism} \label{subsection: free_will}

Although the efficacy of brain- and behaviour-based prediction on lie and intention needs to be improved, perhaps many will agree with the existence of lie and intention, their neural correlates, and their implications in law. The existence of the free will, however, has not been universally agreed upon \citep{hume1748enquiry, nichols2011, heisenberg2009free,  soon2013predicting, haynes2011decoding, bode2011tracking, soon2008unconscious, haynes2007reading}. In brief, free will states that one’s brain (or mind) decides on its own rather than being governed by some deterministic rule. It nevertheless has a profound implication on the justice system. 

My view on free will is an integrated one. On the one hand, I believe that free will exists in higher-order brain functions, including, for example, the PFC’s involvement in moral reasoning \citep{greene2001fmri, heekeren2003fmri, goodenough2001mapping, moll2003morals}. On the other hand, I believe there also exists potentially deterministic operations in lower-order brain functions, such as colour and face recognition \citep{zeki2020bayesian}. Other mental or cognitive faculties and their behavioural manifestations are perhaps generated by integrating the two; namely, they combine some deterministic neural processes and free will. For example, associating a white flag with surrender (combing deterministic colour recognition and the concept of surrender which is free will) and racial profiling (combining deterministic face recognition and prejudice which is arguably free will). 

To promote discussion, suppose free will exists \footnote{In the eyes of the law in the UK, it is considered that people are generally considered to have free will.}. I proceed to discuss its impact on law and decision-making. Research has demonstrated the possibility of predicting free will via brain decoding \citep{wisniewski2019free, haynes2011decoding}. Yet it remains to distinguish criminal behaviour due to free will from criminal behaviour due to brain damage (which disrupts free will). The former would lead to more severe punishment. Relatedly, there is a possibility where free will is lurking in a criminal case, for example, when the PFC of the defendant is damaged. More specifically, some PFC damage may change behaviour significantly, resulting in violent and criminal behaviour \citep{brower2001neuropsychiatry}. But proving PFC damage in court does not necessarily exempt one from committing crimes. For example, minor PFC damage may lead to the inability to feel emotional pain. Between not feeling emotional pain and committing a crime potentially lurks the free will. More precisely, the minor brain damage leading to a lack of emotion (which is not necessarily crime-causing) and free will, which acts upon or in association with the lack of emotion, results in criminal behaviour. Predictive modelling is potentially helpful for assessing whether one ‘lacks capacity'. If one commits a criminal act and evidence is provided (via brain decoding) that they lacked capacity at the time of the offense, then this can be a mitigating factor when sentencing them or considering how to rehabilitate them \citep{Graham2019navigating}. 

The Sheila Berry case \citep{commonwealth2014berry} is perhaps an example where free will was eliminated, so to speak, by mental illness, brain damage, and brain tumor. More specfically, although ``evidence of mental impairment did not warrant reduction of conviction of murder in first degree on theory of extreme atrocity or cruelty'' \citep{commonwealth2000rosenthal}, Berry's lasting mental illness in addition to brain damage and brain tumor prevents her free will to act in full capacity, and her initial twice conviction of murder in the first degree was, therefore, reduced to murder in the second degree. Said in another way, ``...  the defendant's long-standing bipolar or schizoaffective disorder combined with the tumor on her cerebellum were very closely intertwined with her conduct on August 14, 2002, and specifically with her actions in killing the victim ... In particular, ... these conditions affected the defendant's conduct in a manner that has an impact on the determination whether the defendant herself committed the murder with extreme atrocity or cruelty'' \citep{commonwealth2014berry}.

One cannot discuss free without speaking about automatism. Automatism, in simple terms, means that an accused is unaware (\textit{i.e.}, does not have conscious control) of one's actions leading towards illegal actions. In other words, the decision one makes is not (free) willed or deliberate. There are several variants of automatism. If automatism is not due to mental illness, one may obtain a complete acquittal. When automatism is caused by a mental illness, the outcome is a verdict of Not Criminally Responsible on Account of Mental Disorder (NCRMD)  \footnote{``No person is criminally responsible for an act committed or an omission made while suffering from a mental disorder that rendered the person incapable of appreciating the nature and quality of the act or omission or of knowing that it was wrong.'' Section 16 of the Criminal Code, R.S.C. 1985, c. C-46.}  If automatism is due, in part, to intoxication/inebriation, then one could argue for a partial defense to obtain conviction for a less serious included offense \citep{simmons2017free}, but one needs to distinguish voluntary and involuntary intoxication. Whereas it is well known that voluntary intoxication is not automatism, it remains to determine involuntary intoxication and to evaluate whether this leads to exculpation. A recent study suggests a six-step procedure to determine involuntary intoxication \citep{brooks2015involuntary}. Such a ``sequential'' guidance provide a reasonable foundation to translate it into algorithms to potentially automate/fascinate decisions. 

Related to automatism is provocation, where one loses control suddenly due to verbal or physical actions. Although provocation does not constitute automatism, one can use it as a partial defense. Moreover, works have shown that self-control inhabits aggression \citep{dewall2011self, schreck2021predicting}. For example, self-control of adolescence was significant predictors of conflict resolutions \citep{vera2004conflict}.  Self-control was also positively associated with accommodative tendencies to a romantic partner's potentially destructive behavior \citep{finkel2001self}. Finally, individuals with self-control training decreases anger and aggression in response to provocation in aggressive individuals \citep{denson2011self}.

Converging findings suggest that the limbic system (\textit{e.g.}, amygdala), the prefrontal regions, and dorsal and ventral systems are involved regulating emotions and control \citep{kelley2019stimulating}. More specifically, one model \citep{ochsner2007neural} suggests that emotion is modulated via a bottom-up process via the limbic system ((\textit{e.g.}, amygdala) \citep{ledoux1992emotion}) and a top-down appraisal processes that involve prefrontal regions. Another, related model  suggests that emotion is regulated via a cortico-subcortical network consisting in a dorsal system and a ventral system \citep{phillips2003understanding}. 

Combining the neural and legal evidence, one may design a model (\textit{e.g.}, a mediation model \citep{chen2018high, chen2021identifying} where provocation tendency mediates brain functions and criminal actions) to find potential pathways from brain areas to provocation and therefore to potential criminal action; such links may not only explain the neural and behavioural basis of provocation and criminal actions, but also may help design solution to manage or deter anger, aggression, and provocation \citep{denson2011self} and, therefore, potentially reduce criminal actions.

\section{The challenges of applying brain and behaviour decoding in jurisprudence}\label{section:challenges}

\subsection{Privacy, ethics, and prosecutorial abuse} \label{subsection: privacy_ethics_prosecutorial_abuse}

Here, in light of recent development and debates, I discuss a set of concerns machine learning and AI-based legal enquires using neural and behavioural data may face regarding privacy, ethics, and prosecutorial abuses. 

To begin, let’s ask ourselves a few questions. Suppose you are taking a walk in a modern city where your faces have been scanned many times. Who possesses your face data? Who can analyse them? And are you clear to whom the results have been unveiled  \footnote{In recent years, the use of facial recognition data has been heavily regulated. The Information Commissioner’s Office determines what personal data can and cannot be processed and gives out large fines if they consider that a person or entity has breached the GDPR. Companies such as Google, Marriot Hotels and British Airways have also received large fines from GDPR regulators, so there is a real incentive for companies to comply with the GDPR. For example, Amazon was recently fined £636 million for a breach of the GDPR: \url{https://www.bbc.co.uk/news/business-58024116}.} \citep{ico2009use, ico2021use}?
 
Now let’s replace faces with brain images and behaviour measurements and cameras with imaging scanners and smartphones, and think again. Although remote, dynamic brain and behaviour decoding in public may require years’ effort, brain and behaviour decoding, in general, has made notable progress in relatively stable environments, evidenced by works cited in this paper. With data acquisition and analytical methods advancing rapidly, it is, therefore, perhaps timely and necessary to discuss how to prevent a powerful (potentially unjust) “brain and behaviour decoder” from assessing one’s brain signals and behaviour measurements recorded from hospitals, labs, and remotely at home, decoding one’s thoughts and actions, and revealing them to others for undisclosed purposes. 

The brain data can be considered as ‘special category data’ under the General Data Protection Regulation (GDPR) with regulations provided by the GDPR regarding how to process them and under what circumstances \citep{ico2018special}. Nevertheless, “[c]ategoriation of brain-derived data is unclear in terms of the GDPR when it is not about health, nor stemming from medical devices” \citep{rainey2020european}. Perhaps more confusingly, if one day machines develop into becoming capable of “remote brain decoding” outside of laboratories, it is unclear whether they are permitted to record one’s brain signals, whether they are allowed to decode the data (\textit{e.g.}, extraction and extrapolation), and with whom they can share the results. 

As a part of personal data \footnote{The term ‘personal data’ refers to ``any information relating to an identified or identifiable natural person (‘data subject’)''; an identifiable natural person ``can be identified, directly or indirectly, in particular by reference to an identifier such as a name, an identification number, location data, an online identifier or to one or more factors specific to the physical, physiological, genetic, mental, economic, cultural or social identity of that natural person'' (see GDPR Article 4).} \footnote{See also GDPR Article 13: The subject needs to receive “meaningful information about the logic involved” in automated processing, and GDPR Article 22: the data controller needs to implement measures to ensure the data subject’s rights, freedoms, and legitimate interests, for example, to contest the decision.} that constitute of, and are largely \footnote{There are shared functional and structural characteristics between brains, such as those in the primary and associated visual cortex.} unique to an individual, brain data should, in my view, be protected in the same way identifiable health records \footnote{For example, ``[d]ata from BCIs and other brain recordings is often personal and may be as sensitive as health data'' \citep{rainey2020european}.} (such as disease status and severity) are protected. Indeed, recording brain data in hospitals and research laboratories at present requires strict ethical approval. Yet, there are some untravelled territories - where scholars working on AI/machine learning, brain and behavioural sciences, and law could potentially jointly lead the exploration and discussion. For example, can a prosecutor acquire (or demand to acquire) one’s brain images freely and use them as court evidence (if so, under what circumstances; if not always, when not)? How can we prevent someone from abusing brain and behaviour data analysis to prosecute one's opponent(s) and defend one's client(s)? 

Unlike personal information or digital records, which are relatively concrete \footnote{One’s health and disease records are either written in physical forms or stored in digital forms; once recorded, they do not usually change significantly.}, and encode information within a comparatively narrow and static scope \footnote{Such as age and gender; they do not change as widely and wildly as individual (brain) thoughts.}, brain data are dynamic, variable, and contain limitless variability. By limitless, I mean there are perhaps a variety of ridiculous thoughts that may be going through our brain (which do not necessarily end up in verbal or behavioural outputs) from moment to moment. If one day we can relatively transparently decode the brain, these thoughts may become unnecessarily apparent to others in court. Imagine a ``brain decoder'' reveals that one thinks the jurors dress funny or are being unfair; would that not affect some of the jurors emotionally? Similarly, all stored memories, ridiculous or sensitive, are also at risk for others to see and potentially bias jurors’ views on the individual, in a way perhaps similar to jurors' racial biases \citep{korn2012neurolaw}. 

At the extreme end, suppose there is one person who frequently thinks about committing (but does not practice) homicide. Decoding thoughts may jeopardise one's life and career and, due to stress and unwanted attention, may push him/her towards committing wrongdoings or crimes later in action rather than thoughts. Certainly, one could argue this person may have an underlying psychological or mental substrate and may be more likely (probabilistically speaking) to render crimes in the future, and therefore needs to be managed or treated; but this person currently has not committed anything unlawful and should not be, per law, punished. 

Taken together, I argue that unless having the brain data available to the court will add additional, unbiased information towards making a better judgment and decision-making, one’s brain data should be treated as strictly as one’s private health data (such as data recorded from health apps on smartphones) following the GDPR, if not stricter (for reasons above). Additionally, further discussions are needed to clarify circumstances under which one’s brain (and behaviour) signals can be analysed to extract evidence without the individual’s, or, in cases where one cannot give one's consent (\textit{e.g.}, one has psychological and psychiatric illnesses), one's family’s or physicians’ consent. Equally, consensuses need to be reached regarding situations where one's brain (and behaviour) signals cannot be analysed, especially to clarify under what scenarios there needs strict prohibition or protection. A stringent guideline may protect cases where the brain and behavioural analyses may add little legal insights but cause one’s irregular brain activities to be unnecessarily revealed in the court (even just to one's family and friends), potentially bringing stigma or prejudice to the individual's daily life afterwards. Finally, although there have been clear regulations \citep{coucil1999convention} on when predictive tests can be made using genetic information, there have not been general regulations regarding the applications of predictive tests using brain and behaviour data, particularly in legal cases. There is, therefore, an urgent need to clarify the definitions, set the boundaries, and establish regulations \citep{chesterman2021we} regarding when and when not brain and behaviour data can be used in a predictive manner in legal cases.

\subsection{Accuracy, reliability, and reproducibility} \label{subsection: accuracy_reliability_reproducibility}
To date, brain- and behaviour-based predictions are largely based on models developed from laboratory settings where the conditions are controlled and noises reduced. Models and parameters developed under these circumstances may not reflect a broad range of sceneries in the real world or be generalisable to capture the wider variability in a population (especially concerning heavy-tail events \footnote{Events on a heavy-tail distribution very rarely happen (but do happen); as such, algorithms trained from samples may not capture those rare events. For example, suppose a neural lie detector was trained - using true and false statements as well as brain data - when giving these statements to healthy subjects and patients who have known-neurological disorders (such as psychosis) while having their brain data recorded. But suppose there is a subject with a unique type of brain lesion; this person does not belong to the healthy group or the psychosis group; as such, the algorithm may detect this person as lying given the brain data – even if one is not lying. }). Such models, when applied to out-of-the-laboratory features and real-world legal cases, may become irreproducible. For example, in a lab, even when asked to consider countermeasures, a matched control may not strive to avoid a positive detection since no severer consequences are involved. But if a defendant is asked to make a statement (based on which one may receive a death sentence) and simultaneously has, knowingly, one's brain signals recorded, one may make a vigorous mental effort to conceal the truth, such that the brain patterns are partially altered.

Recent research suggests that to robustly identify correlations with individual differences measures, fMRI studies will need thousands (if not tens of thousands) of participants \citep{marek2022reproducible}. This implies that effect sizes concerning brain measures and variation in behaviour, personality \textit{etc}. will be very small. When applied to legal cases, despite promises, the effect sizes - in machine learning-based predictive modelling in general and fMRI-based studies in particular - may be an order of magnitude more challenging than one anticipates.

\subsection{Predicting future actions} \label{subsection: predicting_future_actions}

Using brain imaging techniques, it is possible to show that someone’s brain states are \textit{periodically} disrupted \citep{reinen2018human}. But here I argue that demonstration of \textit{mens rea} via brain decoding at present is perhaps restricted to a small group of individuals whose brain patterns are \textit{continuously} disrupted/irregular. This is, in part, because of the difficulty of linking brain decoding to \textit{actus reus} (criminal conduct). 

More specifically, to convict someone of defence, it must be proved that the defendant had \textit{mens rea} at the precise time of committing the \textit{actus reus}. For example, suppose A wanted to kill B, and A was driving to B’s house with a firearm with the intention of shooting and killing B when A got there. But on the way, a cyclist swerved in front of A’s car (no fault of A’s) and A tried but could not stop in time, and the cyclist was hit by the car and died. By chance, the cyclist was B. Even though A wanted to kill B and was driving over to his house intending to do so; he did not have the intention to hit the cyclist. In this case, A would not be guilty of murder because he did not have the intent to hit and kill the cyclist at the time A hit the cyclist. But A would have been guilty if A noticed B was cycling near A’s car and A intentionally hit and killed B. 

The point is that one has to have the intention to commit the precise act at the same time as one commits that act. It is unlikely (unless the person wears a wearable EEG daily; but see scanned prisoners in New Mexico, USA with ``portable'' MRI machines \textit{in situ} \citep{miller2008investigating}) that we can scan the person’s brain at the precise time of committing the act – typically only after the act. It is under investigation (\textit{e.g.}, research done in the Haynes lab) the possibility to decode whether one had been to a scene (in essence, it is to decode past memory); it is, as far as I am aware, currently difficult to precisely decode (stored) memory. As such, even if brain decoding can show that one with psychosis has brain patterns that are periodically disrupted \citep{reinen2018human}; one cannot ascertain whether the intention of an act happened during the exact period. The data, however, may show a propensity to have a particular mental state; yet I highlight that this is not the same thing as proof of a particular mental state at a particular time. Nonetheless, a stronger argument can perhaps be made if brain decoding can reliably show that the subject’s brain patterns are \textit{continuously} disrupted, or whose disruption had surely covered the period during which the act was carried out.

From a cross-sectional - when (future) time is not concerned - point of view (see the first two panels in \textbf{Figure} \ref{Prediction} (d)), thinking about a crime and carrying out the crime are considered and punished differently. Yet, when time is taken into consideration (see the last panel in \textbf{Figure} \ref{Prediction} (d)), there appears a what I call, for lack of a better words, \textit{transitional state}, where one thinks about a crime, makes plans for it, but has not carried out the crime. More complicatedly, such transitional states may have different properties. An example of a \textit{transitional state of interest} where one thinks about mass-killing, talks about it and posts the idea of it and begins to purchase fire arms, but has not made any attempt. From a legal point, this is a preparatory offense, which means, in English law, ``an act … more than merely preparatory to the commission of an offence'' \citep{stark2018preparatory} \footnote{Criminal Attempts Act 1981, s. 1(1).} \footnote{In German law, this indicates that the taking of ``steps which will immediately lead to the completion of the offense as anticipated by'' the defendant. German Criminal Code, s. 22 (Strafgesetzbuch). This test has been interpreted to mean that the defendant must take a step which, in one's eyes, means that the completed crime will occur without any further substantial intermediate steps \citep{stark2018preparatory}. }. 

Punishing a preparatory act is, in general, legitimate as the steps undertaken may lead to sufficiently concrete danger to any legally-protected interest. Yet to classify it (and to determine the severity of it) is not straight-forward and may involve various of rules each of which requires a threshold. More concretely, to properly classify and determine the degree of preparatory offense, one needs potentially multiple threshold holds - and one has made a preparatory offense if a defendant has passed one or several of such thresholds. Examples of thresholds are, for how long one has been think about the crime and making plans for it, respectively, what types of plans one is making, to what degree is one making the plans (\textit{e.g.}, talking about it with friends, posting it on social media, buying fire weapons from a store), etc.

To make prediction more complex is the different types of transitional predatory offenses. In my view, there are, in general, four classes of transitional predatory offenses. The first class of transitional state is a \textit{transient} transitional state, for example, that one thinks about mass-killing, purchases one weapon, and made no more purchasing (or returned the weapon or give it to others) - it is transient because it only happens during a short period of time and never resumes. A second class is \textit{stable} transitional state: using the same example, one keeps thinking about mass-killing, and continues talking about it and purchasing more weapons. It is stable as one keeps the thoughts and preparations consistently. A third class is \textit{progressive} transitional state where one thinks about mass-killing, talks about it in increasing frequency and purchasing weapons in increasing volume and power. It is progressive as the degree of crime preparation becomes more and more severe. A fourth class is a \textit{dynamic} transitional state, where one thinks about mass-killing, talks about it and purchases a weapon, then stops doing so for a prolonged period of time, then resumes the talking and purchasing, and so on. It is dynamic because the one goes in and out of this preparatory phase with likelihood of committing future crimes going up and down in dynamic form over time. Longitudinal models in general and those considers dynamic systems in particular are useful, but needs further adaptation to crime data and scenarios, to better distinguish, and map different types of course of transitional predatory offenses.

\subsection{One model fits all \textit{vs.} personalised prediction }
\label{subsection:population_vs_personalized}
I have, in most of this paper, assumed a population model for brain- and behaviour-decoding. In other words, we train a predictive model using both brain and/or behaviour data as inputs and labels (such as intention) as outputs from, say, $1,000$ individuals, and we say that this model can be generalized to other, previously unseen subjects. A population model (if proven generalisable and reproducible) is helpful because it extracts, at the population level, the general relationship between brain/behaviour data and the outcomes of legal interest and, as such, can be extended to other, new individuals to make an assessment when labels (outcomes) are not available or observable. 

Indeed, perhaps most would agree that there are shared population-level similarities between individuals (for both the functions and organisation of the brain and for types of crimes); it is nonetheless unreasonable to assume that a population-level model would suit every person. The reason is twofold. \textbf{First}, at the source of the data, not everyone’s brain is the same. When developing brain (and behaviour) based prediction, at present, one typically standardises different brain images onto the same template (similarly, one can standardise the behaviour features). While such practice embraces the population elements of biology and has practical conveniences, it may miss important individual traits. \textbf{Second}, when mapping brain (or behaviour) data onto the outcomes of legal interest under a population model, it is assumed that the same pathways (or parameters) exist for every individual. Yet, it is not always true. For example, a population model discovers that, on average, the disruption of areas 1 and 2 leads to individuals having a specific outcome (say, intention to kill). Yet, there may exist alternative pathways (say areas 3 and 4, or areas 1, 2, 3, and 4) that also lead to the same outcome. There could be, in addition to the population map, some individual maps that deviate from the population norm but are equally important.

One way to address issues related to “one (population) model fits all” is to develop personalised legal predictive models that account for both population-level shared patterns and individual information. The idea is similar to personalised model development and personalised medicine in the healthcare sector \citep{chen2021personalized}. As such, I do not expand on the technicality; rather, I highlight one potential difficulty: that is, it is difficult to obtain brain and behaviour data for each individual (especially in criminal law where individuals are perhaps not as willing to “donate” their biological data as patients do in health and medicine); equally difficult would be to develop a personalised model for each individual given the high-dimensionality of the parameters and complex combinations of criminal scenarios. A compromise can perhaps be made by developing sub-group models – developing a model for each specific sub-group of individuals (say, one model for homicide and another for burglary), or local-group models -  developing a model for a local population (\textit{e.g.}, \citep{lovins2018validating, boccaccini2017field} in Texas, USA and \citep{hanson2014field} in California, USA). One could additionally include \textit{ad hoc} parameters capturing individual or sub-group characteristics.

\subsection{Predictability \textit{vs.} explainability } \label{subsection: predictability_vs_explainability}

Whereas it is important to continue to improve the accuracy of “black-box” models \footnote{Models that take in brain and behaviour data and produce labelled outcomes such as predicted intention, but the intermediate modelling steps are obfuscated or difficult to explain.} that yield accurate prediction (say that of intention), to convince the judges, juries and perhaps the public to adopt these models in legal investigations, it is also important to make them explainable.

Predictive models, however, are sometimes built at the cost of explainability. For example, regularisation methods reduce estimation variance but introduce bias, making the model less explainable. Ensemble methods improve overall predictability by averaging predictions from individual models; meanwhile, the ensembles become difficult to explain \citep{shmueli2010explain}. Neural networks may uncover hidden associations between features and outcomes and yield accurate predictions, but most are as-of-yet difficult to explain. 

One potential way moving forward is to combine expertise from machine learning, law, and biological science communities to work on explainable models in legal investigations. Recent years have seen great effort in making AI/machine learning models explainable \citep{arrieta2020explainable}. There are also discussions of explainable machine intelligence (XAI) in specific fields such as medicine \citep{tjoa2020survey}. Yet it remains to extend XAI to the field of criminal law by incorporating insights from lawyers, judges, and juries. This is, in part, because the data to be trained are domain (law) specific, and to make the parameters interpretable, one needs insights from experts in biology, law, and machine learning. 

For example, suppose a group of machine learning experts find the parameters corresponding to one set of brain areas consistently associated with a certain type of intention; they need to, on the one hand, verify these findings with neurobiologists whether that area is indeed associated with intention and perhaps decision making (therefore the model makes biological sense), and on the other hand, consult the findings with law-experts whether such explanations makes legal sense and whether evidence derived as such can be used in court. 

In return, biologists and law experts may inspire machine learning/AI scientists to develop more targeted models and methods driven by specific biological/legal problems. One example would be to have biologists and law-experts help to find and narrow down more concrete legal cases where the relationship between the input and output is relatively clear (for example, to find out cases where a relatively clear classification exists between the sentenced and not sentenced) and the traditional legal classification labour-intensive; one can then develop algorithms to estimate such a relationship in an automated way and use it as a baseline model to which future models could compare, and from which better (more accurate and/or more explainable) models can be developed. Note that training a (even simple) model on clear cases/data does not mean the application of the model is restricted or the approach is not novel. In my view, as long as the model can be deployed to make suitable predictions on new subjects, it is novel, and, if it addresses a prediction problem in a traditionally labour-intensive case, useful. Another example is to apply neurobiological and legal insights to reduce (redundant or less useful) data. For example, suppose, based on domain knowledge, areas outside of the PFC, hippocampus, and the visual cortex are not significantly involved in predicting whether a suspect has been to a crime scene, machine learning/AI scientists can then develop models using data from only the PFC, hippocampus, and the visual cortex or penalise parameters associated with areas outside of these regions to make potentially better assessment/prediction instead of assigning weights across the entire brain \footnote{Although the weights for areas outside of the PFC and visual cortex could be very small, it inevitably hinders explanation. }.  

\subsection{When can brain and behaviour data be used as court evidence for recommendations, sentences, and verdicts?} \label{subsection: when_can_be_used}
At present, there are several hypothetical occasions one may consider using brain- and behaviour-based findings in justice via machine learning.

The \textbf{first} is when the individual is likely a suspect or a potential re-offender, but there is otherwise no telling evidence. By inquiring into the brain and behaviour data, one may gain additional insights into the case. A recent Dutch study shows that it is possible to predict violent reoffending in prisoners and those on probation using predictive algorithms and routinely collected datasets used by criminal justice agencies \citep{fazel2019prediction}. Yet, as argued about, extreme caution needs to be made to guard against misjudgment. 

The \textbf{second} is when there is reasonable evidence suggesting a high likelihood that one may have (or have not) committed a crime, but no verdict has been achieved. With court approval and an individual's (or one's physician's) consent, results from brain decoding can be used to help seek further evidence (if the likelihood for committing a crime is high) or as an additional piece of insight (if the likelihood for not committing a crime is high) (see ``beyond a reasonable doubt'' above).

A \textbf{third} is that once a verdict or sentence has been made, one examines, retrospectively with consent, brain- and behaviour-based evidence (which had not been used in legal proceedings and decision-making) to verify the decision and to test the performance of brain- and behaviour-based predictive algorithms (\textit{e.g.}, had we made decisions using the brain and behaviour data, would we have come to the same conclusion?). An example of this is a retrospective study by Dahle \citep{dahle2006strengths} to predict recidivism probability. This, together with data accumulation (algorithms are less accurate or generalisable if trained on limited data even if the (limited) data are of high quality), would help develop more accurate and reproducible methods to facilitate future scientific, technical, and legal investigations.

\section{Final remarks} \label{sec:final_remarks}

The creativity of the human brain has in the past helped to write, amend, and defend the law. Recent advances in machine learning have not only introduced the concept of brain and behaviour decoding but also provided examples for it. Today, standing on the shoulders of machine learning, brain and behavioural sciences, and law, the analyses of brain and behaviour data have begun to offer a glimpse into one’s mental status, intention, and thoughts and how they may affect one’s decisions and actions.

Nonetheless, brain decoding is still in its infancy; many exciting ideas need to be explored and tested. In this paper, I have discussed three important areas where advancements in machine intelligence and brain and behaviour studies may facilitate criminal law: to examine mental illness, evaluate insanity, and assess behaviour; to detect lies, biases, and visits to crime scenes; to predict recidivism and decode one’s thoughts and intentions.

Despite promises and advances, much work is needed, in the laboratories and, importantly, outside of the laboratories and in courts, to demonstrate and improve the efficacy and rigour of brain- and behaviour-based assessment, prediction, and decision-making in criminal law. Equally important is to formulate ethical, practical, safe, and reproducible procedures and protocols regarding recording, analysing, sharing, and extrapolating individual brain and behaviour information in legal practices. A beginning can perhaps be made by joining efforts from experts in machine learning, brain and behavioural sciences, and law. I hope that the observations, arguments, and hypotheses I have made, potentially controversial, may sprawl further discussions in evidence gathering, ethics, data possession, security, and privacy, and in testing the potential of employing brain and behaviour decoding in the future legal investigations. 

While the study of the brain, machine, and law may offer remarkable insights about how brain properties are linked to behavior and criminal actions and how to detect them, and despite a part of the attempts are to increase the accuracy of criminal classification and prediction \textit{bona fide}, for example to raise prevention and protection - to protect those who may be harmed by criminal actions, and to prevent and deter future wrongful sentences, we must not yield to discrimination. This is because, while I believe criminal thoughts, decisions, and actions are products of the brain - some of such neurobiological substrates have or are intertwined with social, economical, physical, and chemical (\textit{e.g.}, substance use) causes, and many of which are themselves inter-twisted and -layered (\textit{e.g.}, drug use causes brain damage; meanwhile drug use also results in social discrimination which may then lead to depression) and unfortunate. These unfortunate factors are perhaps, at least sometimes in part, the ``root causes'' that adversely affect brain operations that give rise to criminal enterprises. While it is justifiable and understandable to sentence and punish criminals according to law, one must not ignore those sometimes neglected underlying social and environmental factors. Therefore, to advocate, defend, and promote justice, I think the future of predictive law may also look into addressing some of these (non-neural or -behavioural) root causes of the brain and behaviour changes, in addition to the examination of brain and behaviour data. A beginning can perhaps be made by learning insights from causal inference, network analysis, pathway analysis, and social sciences to study the social/enviromental and brain/behaviour effects on causal (and thus predictable) criminal outcomes. 

Finally, the complexity of criminal thoughts, decisions, and actions require broad and deep probes not only include, but also go beyond, machine intelligence, brain and behaviour sciences, and law. Thus, one should continue to improve the accuracy of the prediction as well as the understanding of the neurological and behavioral underpinnings of criminal thoughts and actions through collaborations, interdisciplinary quests, and honest criticisms. At the same time, one should have a hopeful, but balanced view about classification, continuous assessment, and longitudinal forecast: no matter how far we will have advanced in our methods, in understanding of the causes of crime, and in predicting its effect, one cannot hope for an omnipotent algorithm or be overly complacent with human judgment, for who is able to rightly classify, assess, or forecast Jean Valjean? No computer. And perhaps no Man.


\begin{funding}
Non-declared.
\end{funding}

\bibliographystyle{apa}

\bibliography{reference.bib}       

\end{document}